\documentstyle[15lomcon,cite,epsfig]{article}

\newcommand{\RTau}[1]{R_{\tau, \mbox{\tiny #1}}}
\newcommand{\DeltaQCD}[1]{\Delta^{\mbox{\tiny #1}}_{\mbox{\tiny QCD}}}
\newcommand{\ind}[2]{^{\mbox{\scriptsize $#1$}}_{\mbox{\scriptsize #2}}}
\newcommand{\inds}[2]{^{\mbox{\scriptsize $#1$}}_{\mbox{\tiny #2}}}
\def\Vud{V_{\mbox{\scriptsize ud}}}
\def\Sew{S_{\mbox{\tiny EW}}}
\def\dpew{\delta'_{\mbox{\tiny EW}}}
\def\MTau{M_{\tau}}
\def\Nc{N_{\mbox{\scriptsize c}}}
\def\nf{n_{\mbox{\scriptsize f}}}

\def\va{_{\mbox{\tiny V/A}}}
\def\zva{\zeta\va}

\begin{document}

\pagestyle{plain}
\thispagestyle{empty}

\title{HADRONIZATION EFFECTS IN INCLUSIVE {\Large$\tau$}~DECAY}

\author{A.V.~Nesterenko \footnote{e-mail: nesterav@theor.jinr.ru}}

\address{BLTPh JINR, Dubna, 141980, Russian Federation}

\maketitle\abstracts{ It is shown that the nonperturbative effects due to
hadronization play a crucial role in low--energy strong interaction
processes. Specifically, such effects impose a stringent constraint on the
infrared behavior of the Adler function and play an essential role in the
theoretical analysis of inclusive $\tau$~lepton decay.}

\renewcommand{\thefootnote}{\arabic{footnote}}
\addtocounter{footnote}{-1}

\vspace*{-2.5mm}
\section{Introduction}
\label{Sect:Intro}
\vspace*{-2.5mm}

This paper briefly presents the results of the studies of effects due to
hadronization in the theoretical description of inclusive $\tau$~lepton
decay~\cite{PRD64,MAPT1,MAPT2,NPQCD07,MAPT3,MAPT4,NPQCD11,Prep}. In this
strong interaction process the experimentally measurable quantity
is~\cite{ALEPH05,ALEPH06}
\begin{equation}
\label{RTauGen}
R_{\tau} = \frac{\Gamma(\tau^{-} \to \mbox{hadrons}^{-}\, \nu_{\tau})}
{\Gamma(\tau^{-} \to e^{-}\, \bar\nu_{e}\, \nu_{\tau})} =
\RTau{V}^{\mbox{\tiny $J$=0}} + \RTau{V}^{\mbox{\tiny $J$=1}} +
\RTau{A}^{\mbox{\tiny $J$=0}} + \RTau{A}^{\mbox{\tiny $J$=1}} + \RTau{S}.
\end{equation}
In what follows we shall restrict ourselves to the consideration of
parts~$\RTau{V}^{\mbox{\tiny $J$=1}}$ and~$\RTau{A}^{\mbox{\tiny $J$=1}}$.
The theoretical prediction for these quantities reads
\begin{equation}
\label{DeltaQCDDef}
\RTau{V/A}^{\mbox{\tiny $J$=1}} \!=\! \frac{\Nc}{2}|\Vud|^2\Sew
\!\bigl(\DeltaQCD{V/A} \!+\! \dpew \bigr),
\quad
\DeltaQCD{V/A} \!=\! 2\!\!\int_{m\va^2}^{\MTau^2}\!\!\!
f\Bigl(\frac{s}{\MTau^2}\Bigr) R^{\mbox{\tiny V/A}}(s)
\frac{d s}{\MTau^2},
\end{equation}
where $\Nc=3$ is the number of colors, $|\Vud| = 0.9738 \pm 0.0005$ is
Cabibbo--Kobayashi--Maskawa matrix element~\cite{PDG2010}, $\Sew = 1.0194
\pm 0.0050$ and $\dpew = 0.0010$ stand for the electroweak corrections
(see Refs.~\cite{BNP,EWF1,EWF2}), and $\DeltaQCD{V/A}$~denotes the QCD
contribution. In Eq.~(\ref{DeltaQCDDef}) $\MTau=1.777\,$GeV is the mass of
$\tau$~lepton~\cite{PDG2010}, $m\va$~stands for the total mass of the
lightest allowed hadronic decay mode of $\tau$~lepton in the corresponding
channel, $f(x) = (1-x)^{2}\,(1+2x)$, and
\begin{equation}
\label{RDef}
R^{\mbox{\tiny V/A}}(s) = \frac{1}{2 \pi i}
\lim_{\varepsilon \to 0_{+}}
\Bigl[\Pi^{\mbox{\tiny V/A}}(s + i \varepsilon) -
\Pi^{\mbox{\tiny V/A}}(s - i \varepsilon)\Bigr]\! =
\frac{1}{\pi}\,\mbox{\normalfont Im}\!\lim_{\varepsilon \to 0_{+}}
\!\Pi^{\mbox{\tiny V/A}}(s + i \varepsilon),
\end{equation}
with~$\Pi^{\mbox{\tiny V/A}}(q^2)$ being the hadronic vacuum polarization
function. The superscripts~``V'' and~``A'' will only be shown when
relevant hereinafter.

In general, it is convenient to perform the theoretical analysis of
inclusive $\tau$~lepton hadronic decay in terms of the Adler
function~\cite{Adler}
\begin{equation}
\label{AdlerDef}
D(Q^2) = - \frac{d\, \Pi(-Q^2)}{d \ln Q^2}, \qquad
Q^2 =-q^2=-s.
\end{equation}
Within perturbative approach the ultraviolet behavior of
$D(Q^2)$~(\ref{AdlerDef}) can be approximated by power series
in the strong running coupling~$\alpha\ind{}{s}(Q^2)$
\begin{equation}
\label{AdlerPert}
D(Q^2) \simeq  D\ind{(\ell)}{pert}(Q^2) = 1 + \sum\nolimits_{j=1}^{\ell}
d_{j} \Bigl[\alpha\ind{(\ell)}{s}(Q^2)\Bigr]^{j},
\qquad Q^2\to\infty.
\end{equation}
In this paper we shall restrict ourselves to the one--loop level
($\ell=1$): $\alpha\ind{(1)}{s}(Q^2)\!= 4\pi/(\beta_{0}\,\ln z)$,
$z=Q^2/\Lambda^2$, $\beta_{0}=11-2\nf/3$, $\Lambda$~is the QCD scale
parameter, $\nf$~denotes the number of active flavors ($\nf=2$ is assumed
in what follows), and~$d_{1}=1/\pi$.

\vspace*{-2.5mm}
\section{Perturbative approach}
\label{Sect:TauPert}
\vspace*{-2.5mm}

In this Section we shall study the massless limit, that implies that the
masses of all final state particles are neglected. In this case, by making
use of definitions~(\ref{RDef}) and~(\ref{AdlerDef}), and additionally
employing Cauchy theorem, the quantity~$\DeltaQCD{}$~(\ref{DeltaQCDDef})
can be represented as
\begin{equation}
\label{DeltaQCDCauchy}
\DeltaQCD{} = \frac{1}{2\pi}
\int_{-\pi}^{\pi}\! D\Bigl(M_{\tau}^{2}\,e^{i\theta}\Bigr)
\!\Bigl(1 + 2e^{i\theta} - 2e^{i3\theta} -e^{i4\theta}\Bigr) d \theta,
\end{equation}
see,~e.g., papers~\cite{BNP,DP,ALEPH06} and references therein. In fact,
the only available option within perturbative approach is to directly use
in the theoretical ex\-pres\-sion for~$\DeltaQCD{}$ the perturbative
approximation of Adler function$\,$\footnote{$\,$Despite of the fact, that
Eq.~(\ref{DeltaQCDCauchy}) is only valid for a ``true physical'' Adler
function $D\inds{}{phys}(Q^2)$, which possesses correct analytic
properties in $Q^2$--variable.}~$D\ind{}{pert}(Q^2)$~(\ref{AdlerPert}),
which contains unphysical singularities at low energies. In this case
Eq.~(\ref{DeltaQCDCauchy}) eventually takes the form (see
Refs.~\cite{NPQCD11,Prep} for the details)
\begin{equation}
\label{DeltaQCDPert}
\Delta\ind{}{pert} = 1 + \frac{4}{\beta_{0}}\int_{0}^{\pi}
\frac{\lambda A_{1}(\theta)+\theta A_{2}(\theta)}{\pi(\lambda^2+\theta^2)}
\,d\theta, \qquad
\lambda = \ln \biggl(\frac{\MTau^2}{\Lambda^2}\biggr),
\end{equation}
where~$A_{1}(\theta) \!=\! 1 \!+\! 2\cos(\theta) \!-\! 2\cos(3\theta)
\!-\! \cos(4\theta)$, $A_{2}(\theta) \!=\! 2\sin(\theta) \!-\!
2\sin(3\theta) \!-\! \sin(4\theta)$.

\begin{figure}[h]
\centering
\includegraphics[width=50mm]{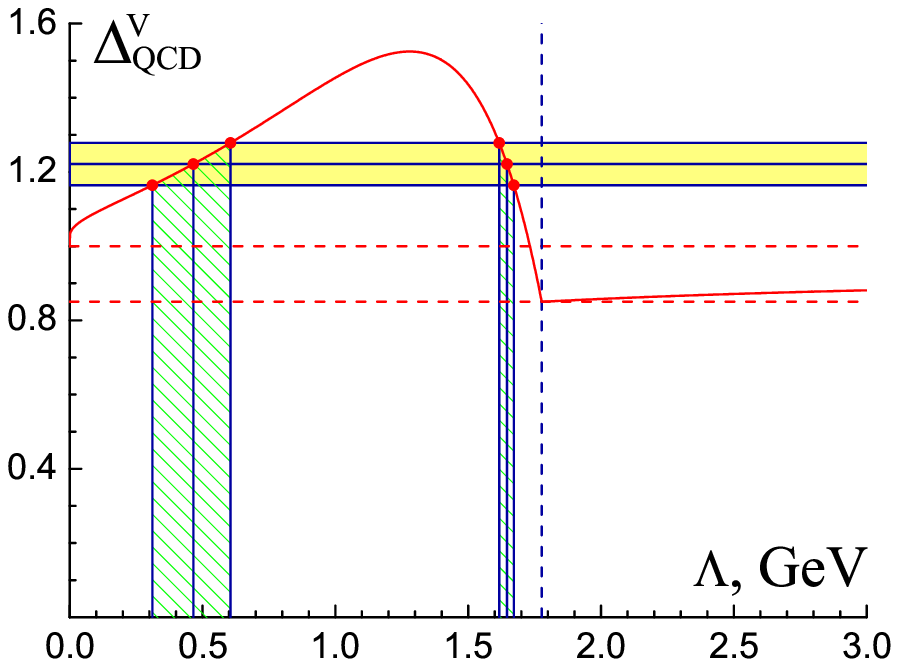}
\hspace{10mm}
\includegraphics[width=50mm]{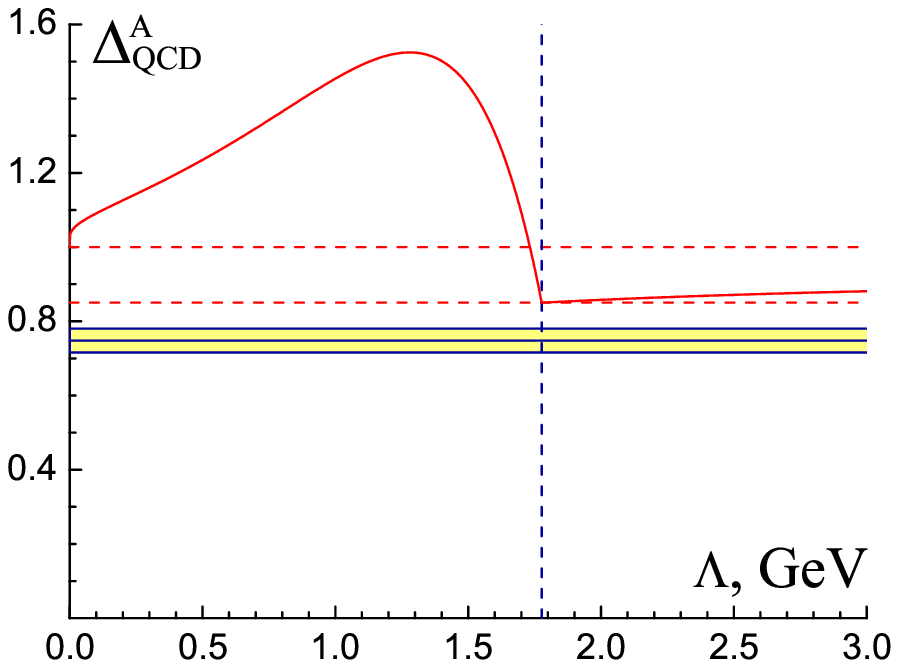}
\caption{Comparison of perturbative expression $\Delta\ind{}{{\tiny
pert}}$~(\ref{DeltaQCDPert}) (solid curves) with experimental
data~(\ref{DeltaQCDExp}). The left and right plots correspond to vector
and axial--vector channels, respectively.}
\label{Plot:Pert}
\end{figure}

It is worth noting here that perturbative approach gives identical
predictions for functions~$\DeltaQCD{V/A}$~(\ref{DeltaQCDDef}) in vector
and axial--vector channels (i.e., $\Delta\ind{\mbox{\tiny V}}{pert} \equiv
\Delta\ind{\mbox{\tiny A}}{pert}$). However, their experimental values are
different, namely~\cite{ALEPH98,ALEPH05,ALEPH06}
\begin{equation}
\label{DeltaQCDExp}
\Delta\ind{\mbox{\tiny V}}{exp} = 1.221 \pm 0.057, \qquad
\Delta\ind{\mbox{\tiny A}}{exp} = 0.748 \pm 0.032.
\end{equation}
For vector channel the comparison of perturbative
result~(\ref{DeltaQCDPert}) with experimental data~(\ref{DeltaQCDExp})
gives $\Lambda\!=\!\bigl(465^{+140}_{-154}\bigr)\,$MeV (the second value,
$\Lambda\!=\!\bigl(1646^{+26}_{-29}\bigr)\,$MeV, will not be considered
herein), see Fig.~\ref{Plot:Pert}. As for the axial--vector channel, the
perturbative approach fails to describe experimental data on $\tau$~lepton
decay.

\vspace*{-2.5mm}
\section{Dispersive approach}
\label{Sect:TauDisp}
\vspace*{-2.5mm}

It is crucial to emphasize that the analysis presented in
Sect.~\ref{Sect:TauPert} entirely leaves out the effects due to
hadronization, which play an important role in the studies of strong
interaction processes at low energies. Such effects were properly
accounted for in the framework of Dispersive approach to~QCD, that has
eventually led to the following integral representations for
functions~(\ref{RDef}) and~(\ref{AdlerDef}) (see
Refs.~\cite{MAPT2,NPQCD07,MAPT4,NPQCD11} for the details):
\begin{equation}
\label{RMAPT}
R(s) = \biggl(1-\frac{m^{2}}{s}\biggr)^{{\!\!}3/2}\!
\!+ \theta\biggl(\!1-\frac{m^2}{s}\biggr)\!
\!\int_{s}^{\infty}\!
\rho(\sigma)\, \frac{d \sigma}{\sigma},
\end{equation}
\begin{equation}
\label{AdlerMAPT}
D(Q^2) = 1 \!+\! \frac{3}{\xi} \!+\!
\frac{3u(\xi)}{2\xi}\,\ln\!\Bigl[1+2\xi\bigl(1-u(\xi)\bigr)\!\Bigr]
\!+\! \frac{1}{u^2(\xi)} \!
\int_{m^2}^{\infty}\!\! \rho(\sigma)\,
\frac{\sigma - m^2}{\sigma+Q^2}\,
\frac{d \sigma}{\sigma}.
\end{equation}
It is worth noting that Eq.~(\ref{RMAPT}) by construction automatically
takes into account the effects due to analytic continuation of spacelike
theoretical results into timelike domain and Eq.~(\ref{AdlerMAPT})
embodies the nonperturbative constraints, which relevant dispersion
relation imposes on the Adler function. In Eqs.~(\ref{RMAPT})
and~(\ref{AdlerMAPT}) $\theta(x)$~denotes the unit step--function,
$u(\xi)=\sqrt{1+1/\xi}$, $\xi=Q^2/m^2$, and~$\rho(\sigma)$ is the
spectral density. For the latter the following expression will be
employed~\cite{NPQCD11}:
\begin{equation}
\label{RhoDef}
\rho(\sigma) = \frac{4}{\beta_{0}}\frac{1}{\ln^{2}(\sigma/\Lambda^2)+\pi^2} +
\frac{\Lambda^2}{\sigma},
\end{equation}
see also Refs.~\cite{PRD64,MAPT1,MAPT4,Rho}. In the right--hand side of
Eq.~(\ref{RhoDef}) the first term is the one--loop perturbative
contribution, whereas the second term represents intrinsically
nonperturbative part of the spectral density. Within the approach on hand
the quantity $\DeltaQCD{V/A}$~(\ref{DeltaQCDDef}) ultimately takes the
following form:
\begin{eqnarray}
\label{DeltaQCD_MAPT_ST}
\DeltaQCD{V/A} \!\!\!\!\!&=&\!\!\!\!\!
\sqrt{1-\zva}\,\biggl(\!1 + 6\zva - \frac{5}{8}\zva^{2}
+\frac{3}{16}\zva^{3}\!\biggr)
\!+\! \int_{m\va^{2}}^{\infty}\!\!H\!\biggl(\frac{\sigma}{M_{\tau}^{2}}\biggr)
\rho(\sigma)\frac{d \sigma}{\sigma} \nonumber \\
&&\!\!\!\!\!
-3\zva \biggl(\!1 + \frac{1}{8}\zva^{2} - \frac{1}{32}\zva^{3}\!\biggr)
\ln\biggl[\frac{2}{\zva}\Bigl(1+\sqrt{1-\zva}\Bigr)-1\biggr],
\end{eqnarray}
where $\zva = m\va^{2}/\MTau^{2}$, $H(x) = g(x)\theta(1-x) +
g(1)\theta(x-1) - g(\zva)$, and $g(x) = x (2- 2x^2 + x^3)$, see
papers~\cite{NPQCD11,Prep} and references therein for the details.

\begin{figure}[t]
\centering
\includegraphics[width=50mm]{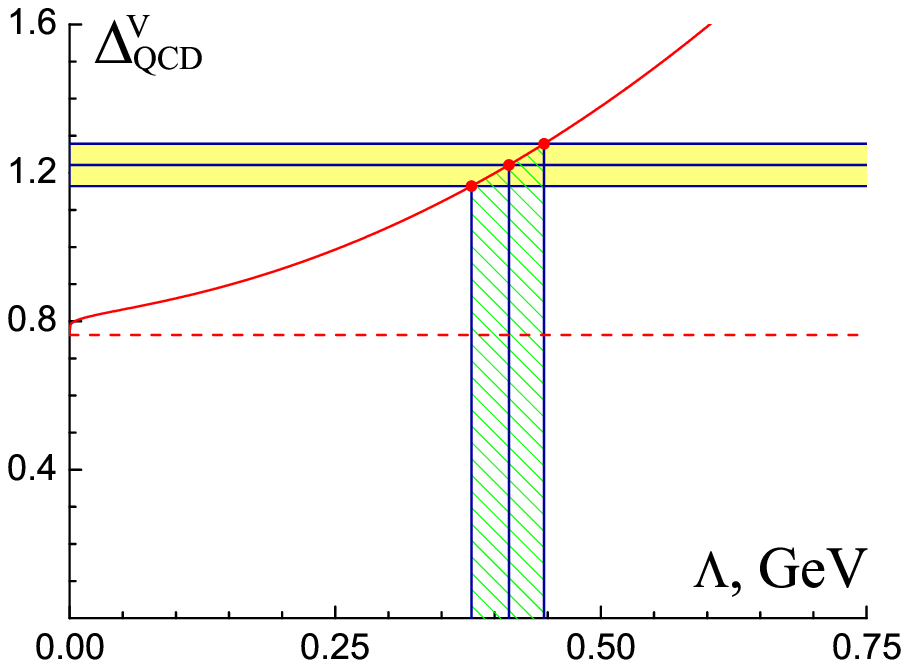}
\hspace{10mm}
\includegraphics[width=50mm]{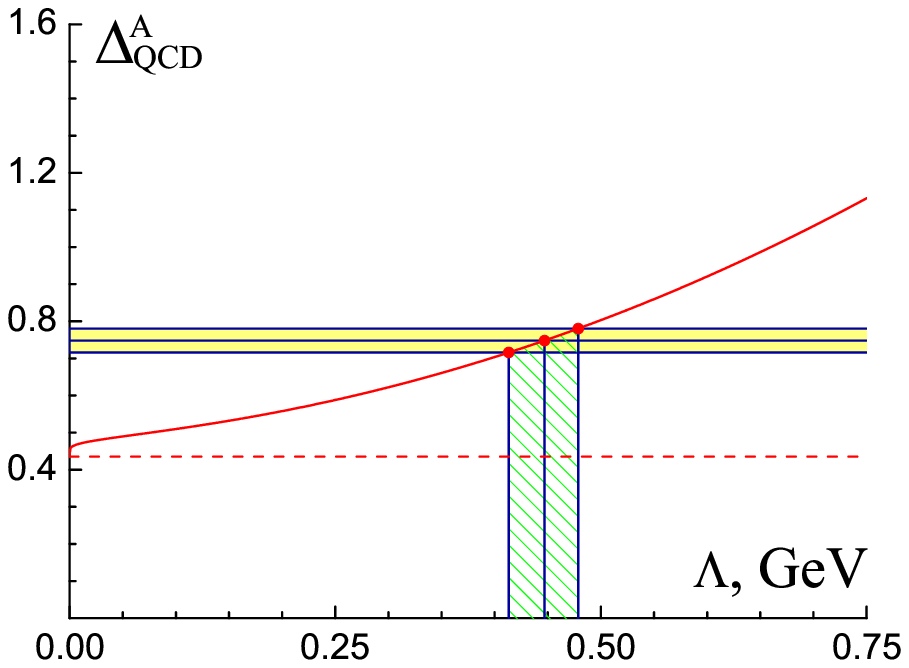}
\caption{Comparison of expression
$\Delta\inds{}{QCD}$~(\ref{DeltaQCD_MAPT_ST}) (solid curves) with
experimental data~(\ref{DeltaQCDExp}). The left and right plots correspond
to vector and axial--vector channels, respectively.}
\label{Plot:MAPT_ST}
\end{figure}

The comparison of obtained result~(\ref{DeltaQCD_MAPT_ST}) with
experimental data~(\ref{DeltaQCDExp}) yields nearly identical values of
QCD scale parameter~$\Lambda$ in both channels, namely, $\Lambda =
\bigl(412 \pm 34 \bigr)\,$MeV for vector channel and $\Lambda = \bigl(446
\pm 33 \bigr)\,$MeV for axial--vector one, see Fig.~\ref{Plot:MAPT_ST}.
Additionally, both these values agree very well with perturbative estimation
of QCD scale parameter described in Sect.~\ref{Sect:TauPert}.

\vspace*{-2.5mm}
\section{Conclusions}
\vspace*{-2.5mm}

The significance of effects due to hadronization in the theoretical
description of inclusive $\tau$~lepton decay is convincingly demonstrated.
The Dispersive approach to QCD has proved to be capable of describing
experimental data on $\tau$~lepton hadronic decay in vector and
axial--vector channels. The vicinity of values of QCD scale
parameter~$\Lambda$ obtained in both channels testifies to the
self--consistency of employed approach.

\end{document}